\documentclass[prb,aps,epsf,twocolumn,showpacs,10pt]{revtex4-1}
\usepackage{graphicx,amsfonts,times,bm,amsmath,verbatim,color,array}

\begin{document}
\title{Normal state Nernst effect from bi-directional bond density wave state in high $T_c$ cuprates}

\author{Girish Sharma$^1$}
\author{Chunxiao Liu$^{2}$}
\author{Kangjun Seo$^3$}
\author{J. D. Sau$^2$}
\author{Sumanta Tewari$^1$}

\affiliation{ $^1$Department of Physics and Astronomy, Clemson University, Clemson, SC 29634\\
$^2$Condensed Matter Theory Center and Joint Quantum Institute, Department of Physics,
University of Maryland, College Park, Maryland 20742-4111\\
$^3$School of Natural Sciences, University of California Merced, Merced, California 95343}

\begin{abstract}
The role of charge order in the phase diagram of high temperature cuprate superconductors has been recently emphasized by the experimental discovery of an incipient bi-directional charge density wave (CDW) phase in the underdoped regime. In a subset of the experiments, the CDW has been found to be accompanied by a $d$-wave intra-unit-cell form factor, indicating modulation of charge density on the oxygen orbitals sandwiched between neighboring Cu atoms on the CuO planes (the so-called bond-density wave (BDW) phase). Here we take a mean field $Q_1=(2\pi/3,0)$ and $Q_2=(0,2\pi/3)$ bi-directional BDW phase with a $d$-wave form factor, which closely resembles the experimentally observed charge ordered states in underdoped cuprates, and calculate the Fermi surface topology and the resulting quasiparticle Nernst coefficient as a function of temperature and doping. We establish that, in the appropriate doping ranges where the low temperature phase (in the absence of superconductivity) is a BDW, the Fermi surface consists of electron and hole pockets, resulting in a low temperature negative Nernst coefficient as observed in experiments.
\end{abstract}

\maketitle
\section{Introduction}
The origin and character of the enigmatic pseudogap phase in the underdoped regime of high temperature cuprate
superconductors remains an open problem \cite{Norman:2005, Keimer}. While the insulating parent compounds of these systems are well understood as three-dimensional (3D) antiferromagnetic Mott insulators, the normal (non-superconducting) phase above superconducting transition temperature ($T_c$) at finite hole doping evinces an anisotropic spectral gap (pseudogap) at low energies below a temperature scale $T^*>T_c$ and behaves strikingly differently from a Fermi liquid. Understanding the pseudogap phase from which superconductivity develops at lower temperatures is generally understood to be the key to understanding the $d$-wave superconducting pairing and the anomalously high transition temperature of the superconducting phase of the high $T_c$ cuprates.
\begin{comment}
As holes are introduced, the superconducting transition temperature ($T_c$) is finite above a critical doping, reaching a maximum at a hole concentration  known as optimal doping. The regime with hole doping less than optimal -- known as underdoped regime -- hosts the remarkable pseudogap phase. In this regime, the normal (non-superconducting) phase above $T_c$ has an anisotropic spectral gap of unknown origin (`pseudo'-gap), and behaves strikingly differently from a Fermi liquid. It is generally understood that the crux of the problem of $d$-wave pairing resides in the pseudogap phase, from which superconductivity develops at lower temperatures.
\end{comment}

 Recent theoretical and experimental work has proposed the role of various charge, spin, electron nematic, and current ordered states competing with superconductivity, and also the role of superconducting fluctuations themselves, to explain the pseudogap phase above $T_c$ ~\cite{Norman:2005,Keimer,Varma:1997,Chakravarty:2001,Kivelson:2003,Chakravarty:2013,Wu:2011,Wu:2013}.
In the class of materials YBa$_2$Cu$_3$O$_{6+x}$ (YBCO), exquisite quantum oscillations of various electronic properties with the applied magnetic field, strong enough to suppress superconducting fluctuations and reveal the underlying normal state, have revealed small Fermi pockets in excitation spectrum reminiscent of a broken symmetry state competing with superconductivity in the under doped regime. \cite{Doiron-Leyraud:2007,Sebastian:2008}. The existence of such small Fermi pockets in the underdoped regime combined with a large hole like Fermi surface in the overdoped regime, indicates that the normal state of the cuprates, in the absence of superconductivity, goes through a Fermi surface reconstruction somewhere near optimal doping. Independent evidence of a similar Fermi surface reconstruction -- from being large and hole-like in the overdoped phase to small and electron-like at underdoping -- is also apparent from the measurements of low temperature Hall and Seebeck coefficients which turn from positive at higher doping to negative in the underdoped regime \cite{LeBoeuf:2007,Chang:2010,Laliberte:2011}. Since the signs of the Hall and Seebeck coefficients are determined by the sign of the dominant charge carriers, the low temperature negative sign of these coefficients in the underdoped regime (in the absence of superconductivity) can be explained by the existence of electron pockets. Interestingly, the low temperature Nernst coefficient, which measures the transverse voltage induced by a longitudinal thermal gradient in the presence of a perpendicular magnetic field, has also been found to be negative in the under doped regime, while being vanishingly small at higher doping. While the sign of the Nernst coefficient, unlike that of Hall and Seebeck coefficients, is not directly related to the sign of the dominant charge carriers, but also depends of the curvature and topology of the Fermi surface, the strikingly different behaviors of the low temperature Nernst response at low and high hole dopings also point to the existence of a Fermi surface reconstruction near optimal doping. 
Although various charge, spin, and current ordered states have been proposed to account for the Fermi surface reconstruction in YBCO \cite{Chakravarty-Kee,Millis-Norman}, none had so far been observed in bulk-sensitive probes until recently.

 In recent x-ray diffraction experiments two groups have independently found strong evidence for a short range charge density wave phase below
 the pseudogap temperature scale $T^*$ for a range of hole doping in the underdoped regime of YBCO \cite{Chang:2012,Ghiringhelli:2012,Kivelson,Tranquada}.
  In these experiments it is not conclusively known if the x-ray diffraction peaks derive from an equal distribution of domains with uni-directional stripe-like correlations  or from correlations with wave vectors $(q_1,0,0.5)$ and $(0,q_2,0.5)$ (with $q_1 \sim q_2 \sim 0.31$) co-existing as in a bi-directional CDW state. However, the lack of anisotropy in the scattering signals  such as intensities and widths, and also salient difference from the stripe like states as observed in the LSCO family such as absence of a coincident magnetic order and a strikingly different behavior of the modulation wave vector with hole doping \cite{Canosa}, indicate that the short-ranged CDW correlations observed in the YBCO family may be different from stripes and in fact an incipient CDW order which is bi-directional. Although the temperature ($T$) dependence of the correlation length above $T_c$ and only short ranged correlations in the CuO$_2$ planes indicate that the observed charge order is only quasi-static, the near divergence of the correlation length as $T \rightarrow T_c$ and that the scattering signals significantly increase on application of magnetic field below $T_c$ indicate that a true thermodynamic CDW transition at some critical temperature ($T_{CDW} < T_c$) may be preempted by the superconducting transition at $T_c$. Furthermore, recent inelastic x-ray scattering  and nuclear magnetic resonance experiments indicate that the short-range charge order observed below $T^*$ is in fact truly static \cite{MTacon,TWu}, presumably due to pinning by disorder potential. Evidence for a similar charge density wave transition in the underdoped regime has also been found in other recent experiments~\cite{LeBoeuf:2013,Hinton:2013,sebastian:2014,Fujita:2014,Comin:2014}. In at least two recent experiments \cite{Fujita:2014,Comin:2014} the charge order has been found to be accompanied by a $d$-wave intra-unit-cell form factor, indicating modulation of charge density on the oxygen orbitals sandwiched between neighboring Cu atoms on the CuO planes (the so-called bond-density wave (BDW) state). %{\tiny {\color{blue}(K.Fujita et al. PNAS 2014, R. Comin et al. arXiv:1402.5415. 2014)}}
Taken together, though it is unclear at the moment if the bond density wave order observed in the cuprates is static and long-ranged or fluctuating and short-ranged  below $T^*$, it is clear that its role in the fermiology of the cuprates should be significant especially at low temperatures ($T< T_c$) and in high magnetic fields (sufficient to suppress superconductivity) where the bi-directional bond density wave is expected to develop long range order resulting in Fermi surface pockets in the single particle spectrum.

An important part of the fermiology of the cuprates is the normal state Nernst effect in the pseudogap phase. The Nernst response, which measures the transverse voltage induced by a longitudinal thermal gradient in the presence of a perpendicular magnetic field, is defined to be positive if dominated by vortices in a superconductor. While the quasiparticle Nernst signal is typically small for conventional metals due to Sondheimer cancellation, the signal carried by vortices can be large and positive in the presence of superconducting fluctuations, as has been found in the cuprates near superconducting $T_c$ and above. Suppressing the superconducting fluctuations by a strong magnetic field reveals the normal state Nernst coefficient ($\nu/T$, with $T$ the temperature) and this has been found to drop with decreasing $T$ in the pseudogap phase, culminating in a negative $\nu/T$ as $T \rightarrow 0$ \cite{Chang-Nernst,Doiron-Nernst}. The low temperature negative Nernst response as $T\rightarrow 0$ is reminiscent of a similar change of sign (with decreasing $T$) in other transport signatures of the pseudogap phase such as Hall and Seebeck coefficients \cite{LeBoeuf:2007,Chang:2010,Laliberte:2011}. The signs of the Hall and Seebeck coefficients are determined by the sign of the dominant charge carriers and can be explained by the existence of an electron pocket centered at $(Q/2, Q/2)$ where the bi-directional BDW state is a superposition of CDWs (with $d$-wave form factors) with ordering wave vectors $(Q,0)$ and $(0,Q)$. This is similar to the recently found result of a change of sign (with decreasing temperature) of the Hall and Seebeck coefficients in the bi-directional CDW state without the $d$-wave form factors \cite{Kangjun}. The sign of the Nernst coefficient, on the other hand, is not directly determined by the sign of the dominant charge carriers and thus may or may not be the same as the sign of the Hall and Seebeck coefficients.

 In this paper we ask if the quasiparticle Nernst coefficient in the mean field BDW state does indeed show a drop with decreasing temperature, with $\nu/T$ eventually becoming negative as $T\rightarrow0$ as seen in experiments. We consider a two-dimensional (2D) bi-directional $Q_1=(2\pi/3,0)$ and $Q_2=(0,2\pi/3)$ BDW state in mean field theory (valid for temperatures $T<T_{BDW}$ and magnetic fields high enough to eliminate the superconductivity) and investigate the quasiparticle Nernst coefficient  as functions of temperature and hole doping appropriate for the underdoped regime of the cuprates. Although the experimental evidence is that for a slight incommensuration in the BDW scattering vectors (i.e., $q_1 \sim q_2 \sim 0.31$) in this paper we work with a commensurate BDW for simplicity (i.e. we take $q_1=q_2=0.33$, corresponding to charge modulations with periodicity of three lattice vectors). We find that, below the BDW transition temperature and in the appropriate regime of hole doping, the Fermi surface topology changes from a large hole-like Fermi surface at higher doping (where there is no BDW) to small Fermi surface pockets at lower doping. A similar fermi surface reconstruction in terms of a CDW state was recently assumed to explain the low frequency of quantum oscillations in the pseudogap phase of the cuprates~\cite{sebastian,sebastian:2014}. We find that the quasiparticle Nernst coefficient in the mean field BDW state does indeed show a drop with decreasing temperature, with $\nu/T$ eventually becoming negative as $T\rightarrow0$, as seen in experiments.
 
 This paper is organized as follows: In Sec. II, we consider the Hamiltonian for the BDW state and examine the energy spectrum and the reconstruction of the Fermi surface. In Sec. III, we define the quasiparticle transport coefficients which we compute numerically using Boltzmann semiclassical equations. Sec. IV and V are devoted to analytical calculations of the Nernst coefficient in the limit of small order parameter and magnetic fields and the breakdown of Sondheimer cancellation. In Sec VI, we present our numerical results for Seebeck, Hall and the Nernst coefficient for the BDW state and show that they all become negative at low temperatures. We end with summary and conclusion in Sec. VII. Some analytic expressions and formulas have been relegated to the appendix.
%An explanation of the Fermi pockets observed in the recent quantum oscillation experiment in Ref.~\cite{sebastian:2014} in terms of the bi-axial CDW state considered in the present work is left for future study.

\section{Model and formalism}

In a mean-field picture, the Hamiltonian describing a density wave ordered state can be written as,
\begin{equation}
H^{DW}=\sum\limits_{\mathbf{k},\mathbf{Q},\sigma}[W(\mathbf{k})c^{\dagger}_{\mathbf{k}+\mathbf{Q},\sigma} c_{\mathbf{k},\sigma} + h.c.],
\end{equation}
where $W(\mathbf{k})$ is the order parameter which can in general describe a charge, orbital current, or a bond density wave in cuprates depending on the form factor $W(\mathbf{k})$. The operator $c^{\dagger}_{\mathbf{k},\sigma}$ creates an electron of spin $\sigma$ with momentum $\mathbf{k}$, and $\mathbf{Q}$ denotes the modulation wave vector. Charge modulations with a periodicity of $1/\delta$ ($1/\delta$ integer) lattice vectors describes a commensurate CDW state. The modulations can be given by a uni-directional modulation $\mathbf{Q_1}=2\pi(\delta,0)$ or $\mathbf{Q_2}=2\pi(0,\delta)$ or a superimposition of the two wave-vectors in which case the CDW is bi-directional. A modulation wave-vector of type $\mathbf{Q}=2\pi(\delta,\delta)$ can describe a third variant of the same CDW state. The functional dependence of the form factor $W(\mathbf{k})$ and the modulation vector $\mathbf{Q}$ distinguish different density wave states for example $\mathbf{Q}=(\pi,\pi)$ and $W(\mathbf{k})=(\cos k_x-\cos k_y)$ is the well known staggered flux or $d$-density wave (DDW) state \cite{Chakravarty:2001}.

%\subsection{Bi-axial bond density wave order}
\begin{figure}
\centering
\includegraphics[width=\columnwidth]{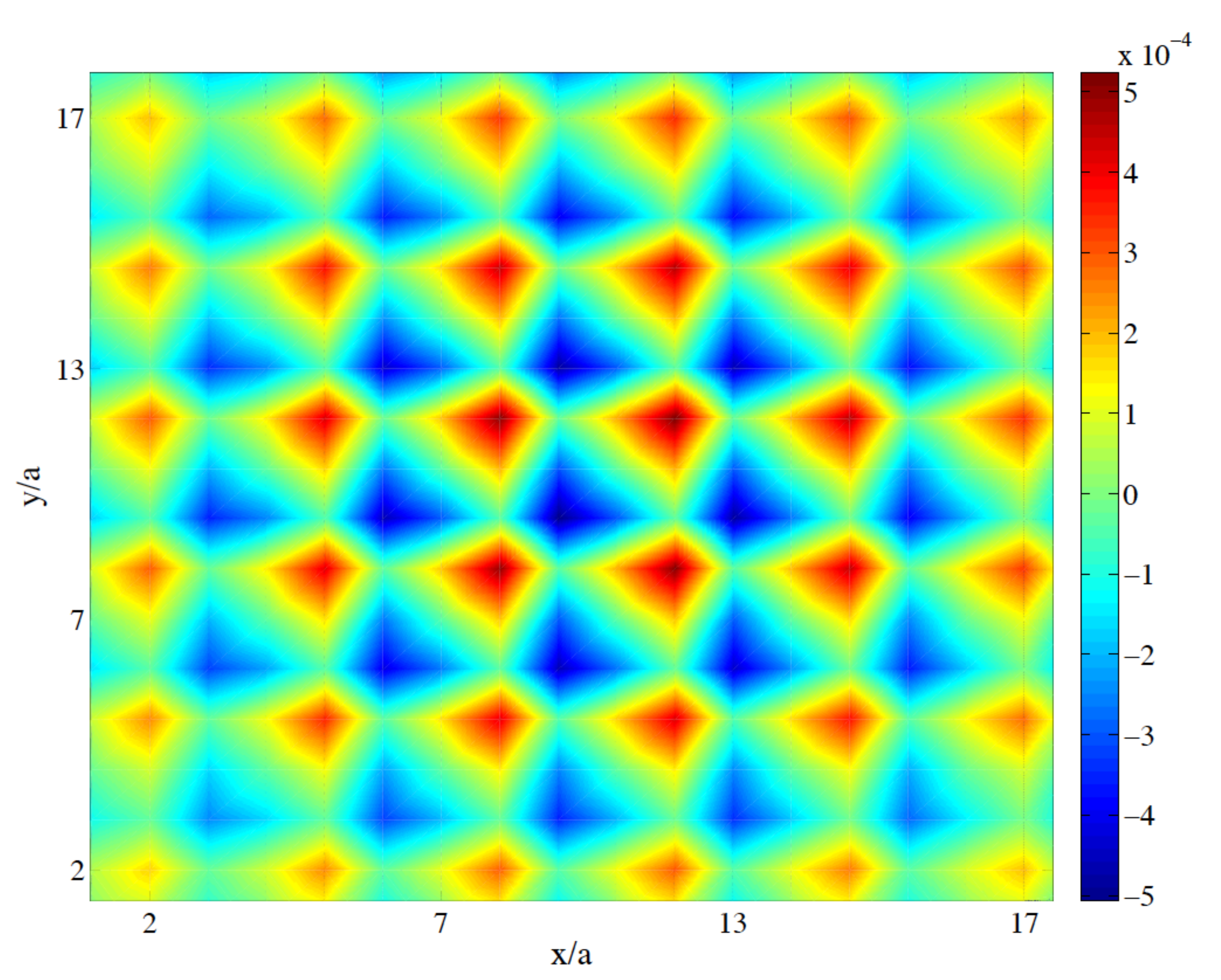}
%\begin{figure}
%\includegraphics[scale=.3]{bdw_realspace_3.pdf}
\caption{(color online) Bond order modulation in real space over a spacing of three lattice vectors. Plot of $|\psi_{BDW}(x,y)|^2 - |\psi(x,y)|^2$ in real space of lattice constant $a$, where $\psi(x,y)$/$\psi_{BDW}(x,y)$ is the lowest energy wave function without/with bond order parameter $W=0.2eV$.}
\label{bdw_realspace}
\end{figure}

\begin{figure}[t]
\includegraphics[scale=.2]{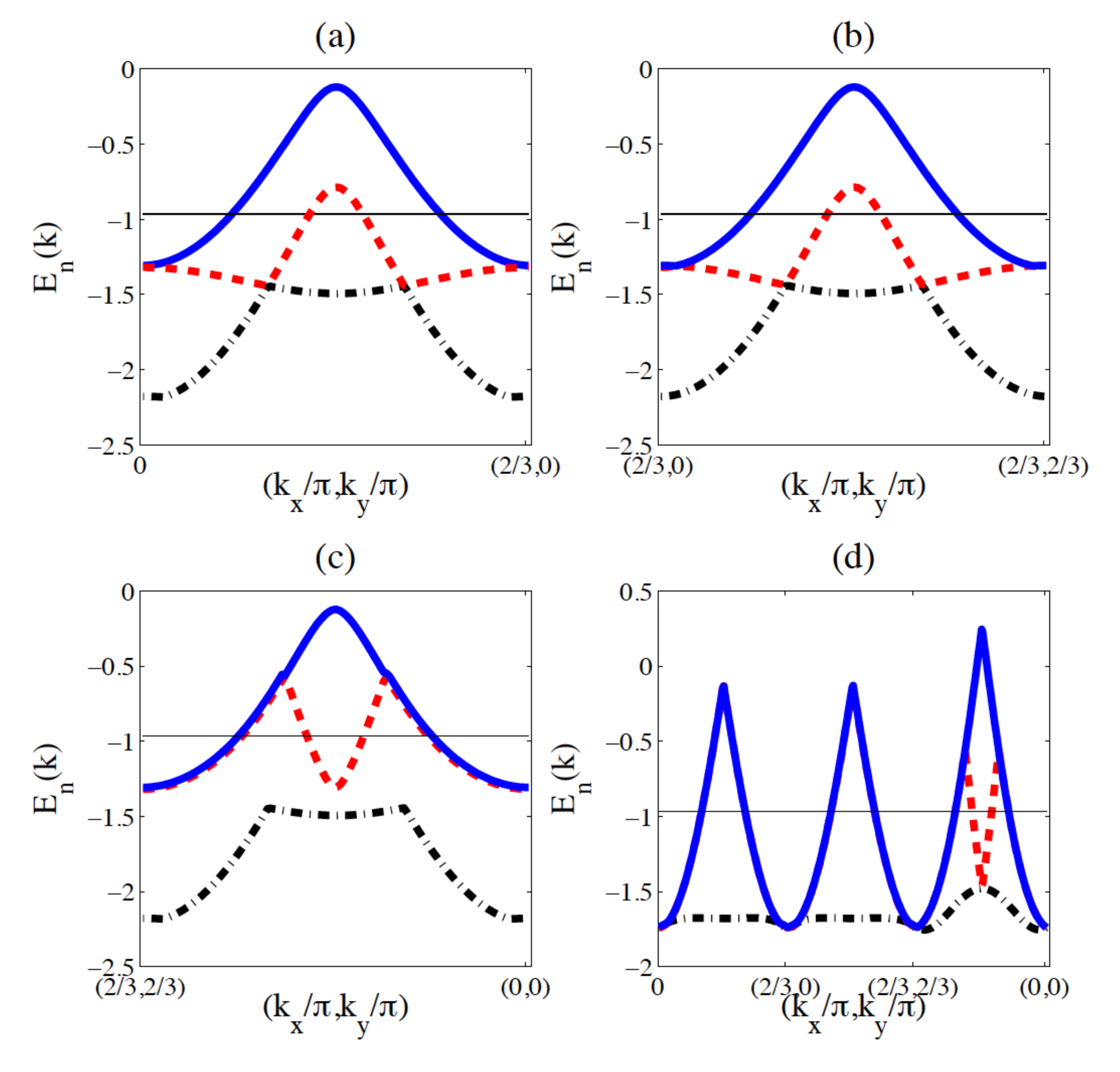}
\caption{(color online) Energy bands of BDW state for $W=0.22 eV$ (a) along the path $(0,0)$ to $(2/3,0)$, (b) along $(2/3,0)$ to $(2/3,2/3)$, (c) from $(2/3,2/3)$ to $(0,0)$ (d) Energy bands in the limit of $W\rightarrow 0$. The solid black line indicates the chemical potential corresponding to a doping of $11\%$.}
\label{EB}
\end{figure}

\begin{figure}[t]
\includegraphics[scale=.2]{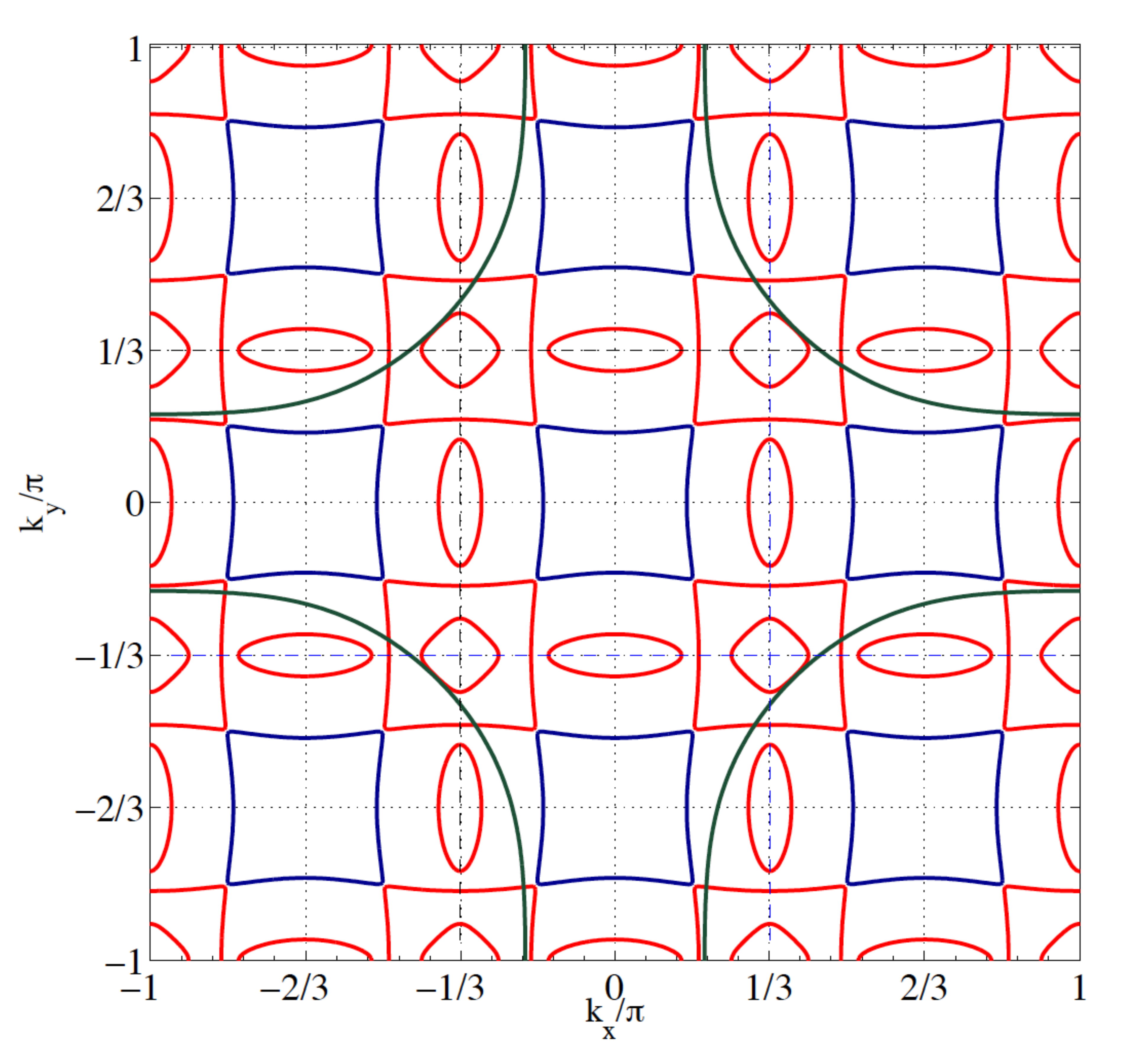}
\caption{(color online) Fermi surface reconstruction for BDW state for $p=11\%$ consisting of an electron pocket at $(\pi/3,\pi/3)$  and hole pockets at $(\pi/3,0)$ and $(0,\pi/3)$. The green contour shows the Fermi surface without the BDW order parameter i.e. $W=0$. The smallest square at the center enclosed by dashed line is the reduced Brillouin zone (RBZ) appropriate for the BDW state.}
\label{FS}
\end{figure}

The tight binding description for electrons on a two-dimensional square lattice of unit lattice constant is given by the energy dispersion relation
\begin{eqnarray}
\epsilon_{\mathbf{k}}&=-2t_1(\cos k_x + \cos k_y ) + 4t_2 \cos k_x \cos k_y \nonumber \\&- 2t_3 (\cos 2k_x  + \cos 2k_y ),
\end{eqnarray}
where $t_1$, $t_2$ and $t_3$ are the nearest neighbor, next-nearest neighbor and next-to-next-neighbor hopping parameters. For all numerical calculations, we chose the parameters $t_1=1.0$ $eV$, $t_2=0.3t_1$ and $t_3=0.1t_2$ which reproduce the non-interacting Fermi surface (see Fig.~\ref{FS}). We now focus on the Hamiltoniain for the BDW state. The following real space mean field Hamiltonian couples fermions to the bond order\cite{Sachdev1}
\begin{eqnarray}
H_{BDW} &= \sum\limits_{\mathbf{r},\mathbf{a},\sigma} [W_{\mathbf{a}} \left(e^{i\mathbf{Q}_1\cdot(\mathbf{r}+\mathbf{a}/2)} + e^{i\mathbf{Q}_2\cdot(\mathbf{r}+\mathbf{a}/2)}\right)c^{\dagger}_{\mathbf{r}+\mathbf{a},\sigma}c_{\mathbf{r},\sigma}\nonumber \\ & + h.c],
\label{BDW1}
\end{eqnarray}
where in the sum $\mathbf{r}$ denotes the lattice sites, the vector $\mathbf{a}$ represents all the nearest neighbors vectors. The operator $c_{\mathbf{r},\sigma}$ annihilates an electron of spin $\sigma$ at the site $\mathbf{r}$. Recent experiments suggest that the bond order $W_{\mathbf{a}}$ resembles the $d$-wave form factor $W_{\pm\hat{x}} = -W_{\pm\hat{y}} = W_0/2$ \cite{Fujita:2014,Comin:2014}. The vectors $Q_1 = 2\pi (\delta,0)$ and $Q_2=2\pi (0,\delta)$ describe the periodic modulation of the bond order where $\delta = 1/3$ indicating a commensurate BDW order with periodicity of three lattice vectors. The bond density wave order effectively redefines the hopping amplitude $t_1$ modulating it spatially. Fig. \ref{bdw_realspace} shows the bond order modulation in real space for the chosen modulation vectors. It is useful to Fourier transform Eq.~(\ref{BDW1}) and rewrite the equation in momentum space:
\begin{eqnarray}
H_{BDW}(\mathbf{k})&=W_0 \sum\limits_{\mathbf{k},\sigma}[ \left(\cos k_x -\cos k_y \right)c^{\dagger}_{\mathbf{k}+{{\mathbf{Q}_1}/{2}},\sigma}c_{\mathbf{k}-{\mathbf{Q}_1}/{2},\sigma} \nonumber\\
&+ c^{\dagger}_{\mathbf{k}+{{\mathbf{Q}_2}/{2}},\sigma}c_{\mathbf{k}-{\mathbf{Q}_2}/{2},\sigma}]+ h.c,
\label{BDW2}
\end{eqnarray}
where $c_{\mathbf{k},\sigma}$ is the annihilation operator for an electron of momentum $\mathbf{k}$ and spin $\sigma$. The total Hamiltonian $H_{MF}$ for the system is
\begin{eqnarray}
H_{MF} = \sum\limits_{\mathbf{k},\sigma} \epsilon_{\mathbf{k}} c^{\dagger}_{\mathbf{k},\sigma}c_{\mathbf{k},\sigma} + H_{BDW},
\end{eqnarray}
which can be expressed in terms of a nine component operator $\Psi_{\mathbf{k},\sigma}$ as
\begin{equation}
H_{MF} = \sum_{\mathbf{k}\in RBZ, \sigma } {\Psi^{\dagger}_{\mathbf{k},\sigma} H(\mathbf{k}) \Psi_{\mathbf{k},\sigma}}
\end{equation}
where $RBZ$ is the reduced Brillouin zone ($-\pi/3<k_x<\pi/3$, $-\pi/3<k_y<\pi/3$) and  $H(\mathbf{k})$ is \begin{widetext}
\begin{equation}
H(\mathbf{k})=\left( \begin{array}{ccccccccc}
\epsilon_{\mathbf{k}} & w_{12} & w_{13} &  w_{14} & 0 & 0 & w_{17} & 0 & 0 \\
w_{21} & \epsilon_{\mathbf{k}+\mathbf{Q}_1} & w_{23} & 0 & w_{25} & 0 & 0 & w_{28} & 0  \\
w_{31} & w_{32} & \epsilon_{\mathbf{k}-\mathbf{Q}_1} & 0 & 0 & w_{36} & 0 & 0 & w_{39}   \\
w_{41} & 0 & 0 & \epsilon_{\mathbf{k}+\mathbf{Q}_2} & w_{45} & w_{46} & w_{47} & 0 & 0   \\
0 & w_{52} & 0 & w_{54} & \epsilon_{\mathbf{k}+\mathbf{Q}_1+\mathbf{Q}_2} & w_{56} & 0 & w_{58} & 0   \\
0 & 0 & w_{63} & w_{64} & w_{65} & \epsilon_{\mathbf{k}-\mathbf{Q}_1+\mathbf{Q}_2} & 0 & 0 & w_{69}   \\
w_{71} & 0 & 0 & w_{74} & 0 & 0 & \epsilon_{\mathbf{k}-\mathbf{Q}_2} & w_{78} & w_{79}   \\
0 & w_{82} & 0 & 0 & w_{85} & 0 & w_{87} & \epsilon_{\mathbf{k}+\mathbf{Q}_1-\mathbf{Q}_2} & w_{89}  \\
0 & 0 & w_{93} & 0 & 0 & w_{96} & w_{97} & w_{98} & \epsilon_{\mathbf{k}-\mathbf{Q}_1-\mathbf{Q}_2}  \\
\end{array} \right)
\label{Hk}
\end{equation}\end{widetext}
Details of the non-zero entries $w_{ij}$ in the Hamiltonian matrix are given in the appendix. Note that hermiticity of the matrix imposes the condition that $w_{ji}=w_{ij}^*$. Diagonalizing the Hamiltonian $H(\mathbf{k})$ in Eq. (\ref{Hk}), we obtain the energy eigenvalues $E_n(\mathbf{k})$ and the corresponding eigenvectors. Fig. \ref{EB} shows the relevant bands of Hamiltonian $H_{MF}$ near the chemical potential out of a total of 9 bands. One notes the presence of an electron pocket centered at $(\pi/3,\pi/3)$ and a hole pocket at $(\pi/3,0)$ and symmetry related points, which are also depicted in the reconstructed Fermi surface in Fig. \ref{FS}. %additionally showing a large electron pocket at $(0,0)$ and a hole pocket at $(0,\pi/3)$. 

\begin{figure}[t]
\includegraphics[scale=.35]{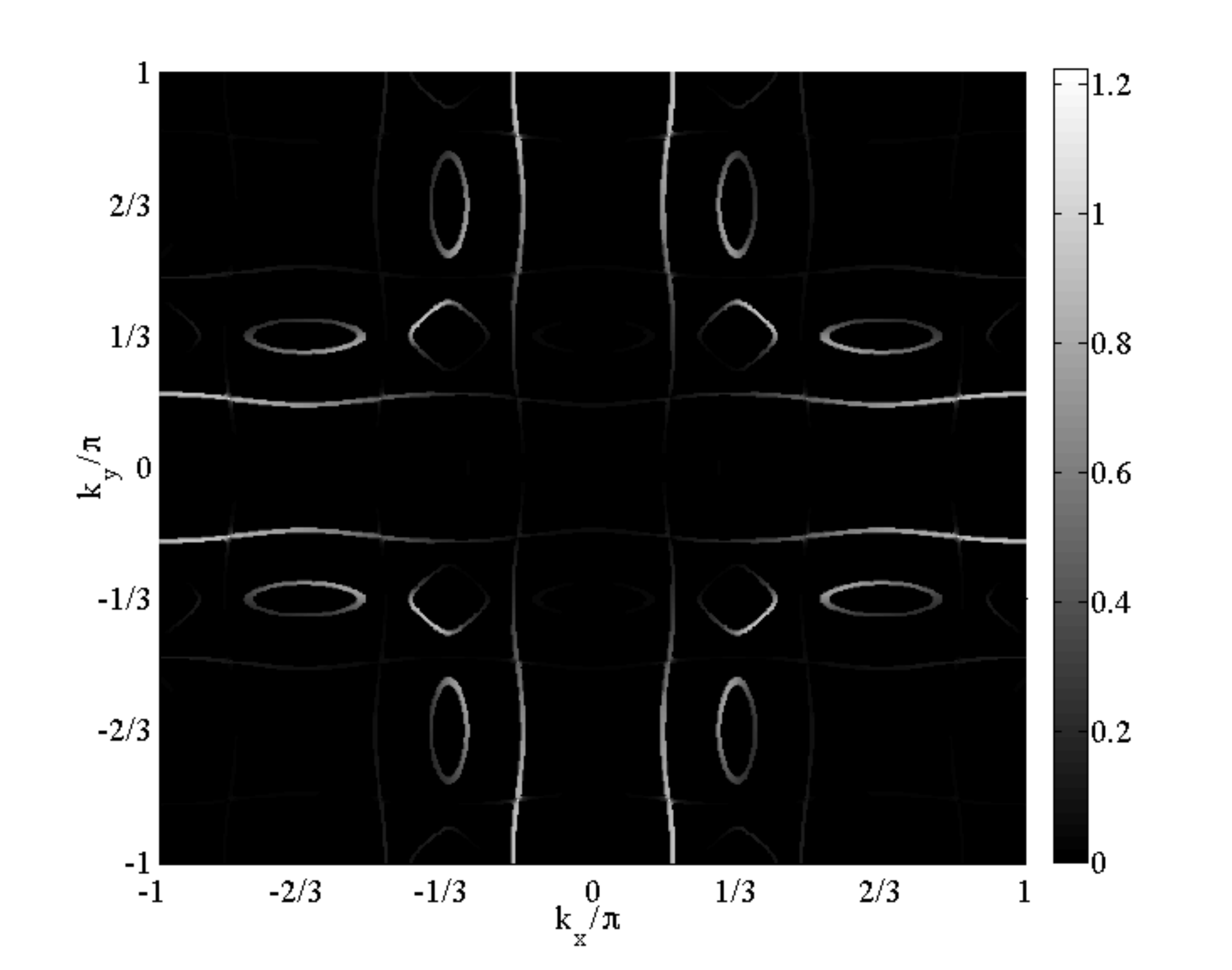}
\caption{Electron spectral function $A(\omega=0, \mathbf{k})$ in the presence of bi-directional bond order $Q_1=(2\pi/3,0)$ and $Q_2=(0,2\pi/3)$ at a doping value of $p=10\%$. Fermi surface reconstruction due to BDW results in the formation of small electron and hole-like pockets which are also observed in Fig. \ref{FS}. The electron spectral function, unlike the bare Fermi surface, is weighted at each point of the BZ by coherence factors and therefore in general is not a periodic function of $Q_1$ or $Q_2$. Similar spectral function for the BDW phase has also been reported earlier \cite{Sachdev1}.}
\label{ARPES}
\end{figure}

In Fig. \ref{ARPES}, we plot the electron spectral function for the BDW Hamiltonian. The electron spectral function $A(\omega,\mathbf{k})$ is given by 
\begin{eqnarray}
A(\omega,\mathbf{k}) = -\frac{1}{\pi}\mbox{Im } G_{\mbox{ret}} (\omega,\mathbf{k}),
\end{eqnarray}
where $G_{\mbox{ret}} (\omega,\mathbf{k})$ is the retarded Green's function for the Hamiltonian. $A(\omega,\mathbf{k})$ essentially maps out the Fermi surface as it should be observed in ARPES experiments. In contrast to the Fermi surface plot in Fig. \ref{FS}, the electron spectral function is not $2\pi/3$ periodic in $k_x$ and $k_y$, but it is weighted by the coherence factors at each point on the Brillouin zone\cite{Sudip1}. A very similar ARPES spectral function for the BDW phase with slight incommensuration has recently appeared in Ref. [\onlinecite{Sachdev1}]. 

The hole doping in the cuprates is conventionally counted from half-filling, i.e., one electron per Cu atom. If $n$ denotes the fraction of occupied number of states in the Brillouin zone then the doping $p=1-2n$. The fraction $n$ is calculated as
\begin{equation}
n=\sum\limits_{n,{\mathbf{k}\in RBZ}}{f(E_n({\mathbf{k}}))},
\end{equation}
where $f(E_n(\mathbf{k})) = 1/(1+e^{\beta(E_n(\mathbf{k})-\mu)})$ is the Fermi distribution function which at zero temperature is simply a step function $\Theta(\mu-E_n(\mathbf{k}))$. We find that $\mu$ behaves linearly with doping $p$. Half filling ($p=0$) is evaluated to be at $\mu=-0.7055 t_1$ and a doping of $p=12.5\% $ is found at $\mu=-1.0016 t_1$.

\begin{comment}
\subsection{Uni-axial bond density wave order}
Here we discuss how the the Fermi surface topology changes due to a uni-axial bond density wave order. Unlike the previous case, here the modulation vector is either $\mathbf{Q}_1=2\pi(0,\delta)$ or $2\pi(\delta,0)$, but again we choose $\delta=1/3$ such that the modulation vector is commensurate with the underlying lattice. The reduced Brilliouin zone now is $0<k_x<\pi/3$ and $-\pi<k_y<\pi$ and analogous to \ref{Hk}, the Hamiltonian $H(\mathbf{k})$ is now a three component matrix

\begin{equation}
H(\mathbf{k})=\left( \begin{array}{ccc}
\epsilon_{\mathbf{k}} & w_{12} & w_{13}  \\
w_{21} & \epsilon_{\mathbf{k}+\mathbf{Q}_1} & w_{23}  \\
w_{31} & w_{32} & \epsilon_{\mathbf{k}-\mathbf{Q}_1}    \\

\end{array} \right) ,
\label{Hk1}
\end{equation}
where the entries $w_{ij}$ in Eq. \ref{Hk1} remain the same as in Eq \ref{Hk} and are given in the Appendix. Again hermiticity imposes the condition $w_{ij}^*=w_{ji}$. Diagonalizing $H(\mathbf{k})$ we obtain the energy eigenvalues $E_n(\mathbf{k})$ which are displayed in figure \ref{eb_2}. There is a hole pocket found at $(\pi/3,\pi)$.

\begin{figure}
\includegraphics[scale=.17]{energybands_uni_1.pdf}
\caption{Plot of three energy bands obtained by diagonalizing matrix $H(\mathbf{k})$ in Eq. \ref{Hk1}. The black line indicates the chemical potential for a doping of $p=12\%$. There is a hole pocket located at $(\pi/3,\pi)$}
\label{eb_2}
\end{figure}

\end{comment}

\section{Quasiparticle transport coefficients}
The sign of the transport coefficients such as the Hall and Seebeck coefficients reveals information about the carrier types (electron or holes) and also the  underlying Fermi surface. We use the formalism of linear response theory to calculate the Hall, Seebeck and Nernst coefficients for the BDW state. The charge current $\mathbf{J}$ and the thermal current $\mathbf{Q}$ can be related to the electric field $\mathbf{E}$ and the temperature gradient $\nabla T$ as
\begin{equation}
\left( \begin{array}{c}
\mathbf{J} \\
\mathbf{Q}  \\
\end{array} \right) =\left( \begin{array}{cc}
\hat{\sigma} & \hat{-\alpha} \\
T\hat{\alpha} & \hat{-\kappa}  \\
\end{array} \right) \left( \begin{array}{c}
\mathbf{E} \\
\mathbf{\nabla T}  \\
\end{array} \right)
\label{thermal}
\end{equation}
The use of three conductivity tensors $\hat{\sigma}$, $\hat{\kappa}$ and $\hat{\alpha}$ is sufficient to relate thermal and electrical effects. The diagonal components of the matrix in Eq \ref{thermal} give us the electrical and thermal conductivity  while $\hat{\alpha}$ interrelates the thermal current and the charge current to electric field and the temperature gradient respectively. Applying a temperature gradient $\nabla T$ across the $x$ axis of the sample, an electric field $E_x$ is generated across given by $E_x=\hat{\sigma}^{-1}_{xx}\hat{\alpha}_{xx}\nabla_x T$ and the Seebeck coefficient $S$ is defined as $S=\alpha_{xx}/\sigma_{xx}$. Applying a magnetic field in the perpendicular direction now generates a Hall current $J_y$ and the Hall coefficient is given by $R_H=\sigma_{xy}/\sigma_{xx}\sigma_{yy}$.

The Nernst effect measures transverse electrical response to a thermal gradient in the absence of a charge current i.e. $E_y=-\vartheta$ $dT/dx$, where $\vartheta$ is the Nernst coefficient and we apply $-dT/dx$ thermal gradient along the $x$ direction which is the appropriate experimental convention. From Eq. \ref{thermal} it follows that the Nernst coefficient $\vartheta$ is
\begin{equation}
{\vartheta} = \frac{E_y}{(-dT/dx)} = \frac{\alpha_{xy}\sigma_{xx} - \alpha_{xx}\sigma_{xy}}{\sigma_{xx}^2 + \sigma_{xy}^2}
\end{equation}
For magnetic field $B$ pointing in the $z$ direction, we redefine the Nernst coefficient to be $\nu=\vartheta/B$. The quantity $\nu/T$ is the one which is determined experimentally. 
 It is important to clarify the sign convention of the Nernst coefficient chosen here according to which the sign of the superconducting Nernst signal is opposite to that of standard textbook convention \cite{Behnia}. According to this convention, the sign of the Nernst signal of the vortices is positive when there is a negative temperature gradient along the $x$ axis.

%\subsection{Bi-axial bond order}
We employ the semi-classical Boltzmann equations approach for the calculation of conductivities in the relaxation time approximation \cite{Allen} with the bi-directional BDW modulation.
\begin{eqnarray}
\alpha_{xx} = \frac{2e}{T}\sum_n\int{\tau(\mathbf{k}) (v_n^x)^2 E_n(\mathbf{k})\frac{\partial f(E_n(\mathbf{k}))}{\partial E_n(\mathbf{k})}d^2\mathbf{k}}
\label{axx}
\end{eqnarray}
\begin{eqnarray}
&\alpha_{xy} = \frac{2e^2B}{T}\sum\limits_n\int{[\tau^2(\mathbf{k}) (v_n^x)^2E_n(\mathbf{k})\frac{\partial f(E_n(\mathbf{k}))}{\partial E_n(\mathbf{k})}}\nonumber\\
&\left(v_n^y v_n^{xy} - v_n^x v_n^{yy}\right)]d^2\mathbf{k}
\label{axy}
\end{eqnarray}
\begin{eqnarray}
\sigma_{xx} = -2e^2\sum_n\int{\tau(\mathbf{k}) (v_n^x)^2 \frac{\partial f(E_n(\mathbf{k}))}{\partial E_n(\mathbf{k})}d^2\mathbf{k}}
\label{sxx}
\end{eqnarray}
\begin{eqnarray}
&\sigma_{xy} = -{2e^3B}\sum\limits_n\int{[\tau^2(\mathbf{k}) v_n^x\frac{\partial f(E_n(\mathbf{k}))}{\partial E_n(\mathbf{k})}}\nonumber\\
&\left(v_n^y v_n^{xy} - v_n^x v_n^{yy}\right)]d^2\mathbf{k},
\label{sxy}
\end{eqnarray}
where $n$ is the band index, $v^x_n$ is the semi-classical quasi-particle velocity $v^x_n=\frac{\partial E_n(\mathbf{k})}{\partial k_x}$ and $v_n^{xy}=\frac{\partial v^y_n}{\partial k_y}$. The integration is restricted to RBZ and the energy eigenvalues $E_n(\mathbf{k})$ are measured relative to the chemical potential. The factor of $2$ present in the numerators takes into account spin-degeneracy of the energy bands and $\tau(\mathbf{k})$ is the scattering time which takes in to account interactions between quasiparticles and impurities, phonons and other quasiparticles. We point out that $\tau(\mathbf{k})$ is assumed to be independent of energy but we retain a possible momentum dependence which yields a positive Seebeck coefficient in the normal state consistent with experiments. We assumed the scattering time $\tau(\mathbf{k})=(1+\alpha(\cos k_x + \cos k_y))^2$, where the parameter $\alpha$ is chosen to be ~0.4  \cite{Kangjun}. 
\begin{comment}
This phenomenological form of $\tau(k)$ is designed to enhance the contribution of the free fermion hole pockets at high temperatures (i.e., above the BDW transition temperature) located near $(\pi,\pi)$ and symmetry related points in the Brillouin zone. By enhancing the contribution of the hole pockets the high temperature ($W=0$) Seebeck coefficient is positive, in accordance with the experiments~\cite{Chang:2012,Ghiringhelli:2012}.  
\end{comment}
The precise functional
form of $\tau(k)$ is unimportant, however, and any other momentum dependence of the scattering time that produces a positive sign of the Seebeck coefficient at high temperatures works just as well. Note that an assumption of a momentum independent $\tau$ results in a negative Seebeck coefficient in the normal state~\cite{hildebrand, kontani, storey} (i.e., above the BDW transition temperature), inconsistent with experiments \cite{LeBoeuf:2007,Chang:2010,Laliberte:2011}. So although the Nernst coefficient $\nu/T$ is robust and negative in the BDW phase as $T\rightarrow 0$ even with a momentum independent $\tau$, which is our central result in this paper, we retain a momentum dependent scattering time only to be consistent with the sign of the high temperature Seebeck coefficient (which is not the focus of this work) \cite{Kangjun}. The temperature dependence of the conductivities arises from the factor of derivative of the Fermi function $\partial f(E(\mathbf{k}))/\partial E(\mathbf{k})$ which takes the form of a Dirac-Delta function at absolute zero.

Note that in our results  we have set $B=\tau({\bf k}=(\pi/2,\pi/2))=1$.
 By examining the definitions of $\alpha$ and $\sigma$, we note that 
we can make $\tau({\bf k})$ and $B$ dimensionless by replacing  $\tau({\bf k})\rightarrow  \tau({\bf k})/\tau_0$ and $B\rightarrow  B(e \tau_0 t_1 a^2/\hbar^2 )$. Here $\tau_0=\tau({\bf k}=(\pi/2,\pi/2))$ 
is a representative scattering time that has been set to $1$ and for our calculations $a\sim 3.9\AA$ is also taken to be 1.
  Choosing $B=1$ to correspond to a physical $B\sim 2T$ (higher values of $B$ does not qualitatively change our results), we obtain that a mean-scattering time of $\tau_0=1$ chosen in this work corresponds to a 
 mean scattering rate $\hbar\tau_0^{-1}\sim 10K$.
% Noting that m_e t a^2/hbar^2 is dimensionless and e B/m_e c = 15K at B=10T; t a^2 m_e/hbar^2=3.83

\section{Analysis of Nernst effect in BDW state using hot-spot model}
\begin{figure}
\includegraphics[scale=.4]{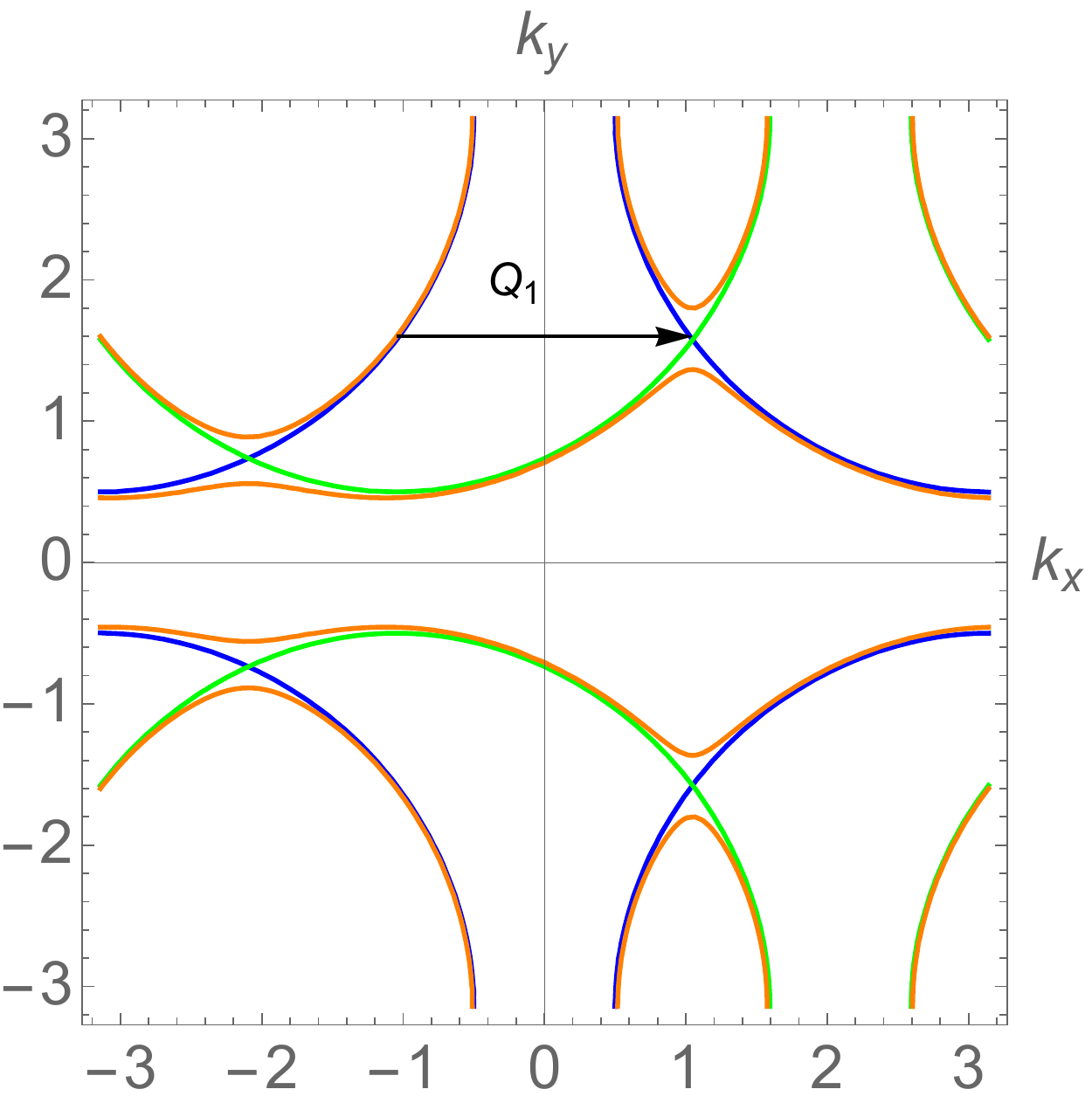}\\
\caption{(color online) The `hot spot' is defined as the point where one piece of unperturbed Fermi surface intersect with another piece when it is translated by the BDW wave-vector $\bm Q_1$. In this figure, blue lines are unperturbed Fermi surfaces, green ones are Fermi surfaces translated by vector $\bm Q_1$, the intersection between blue and green line at the tip of the arrow  is the `hot spot'.}
\label{hotspotfig}
\end{figure}

Before we compute the low temperature Nernst coefficient numerically using Eqs.~\ref{axx} to \ref{sxy}, let us first understand the contribution to Nernst effect from mean field BDW state using the so-called ``hot spot'' model shown in Fig.~\ref{hotspotfig}. 
For weak BDW amplitude $W$, the BDW can only affect electrons at the Fermi surface by scattering by wave-vector $\bm Q_1$. The strongest effect 
of the BDW is felt at momenta $\bm k$, where both the starting wave-vector $\bm k$ and the ending wave-vector $\bm k+\bm Q_1$ are on the Fermi-surface.
The wave-vectors $\bm k$ on the Fermi-surface which satisfy this condition are referred to as hotspots.

 The finite temperature thermoelectric coefficients and conductivities can be written in terms of the zero-temperature conductances as
\begin{align}
\sigma^{T = 0}_{xy}(\mu) & =2 e^3 B\sum_{\pm} \int d \bold{k} \tau^2_{\bold{k}} v^{x}_{\bold{k}} \left[ v^{y}_{\bold{k}} \frac{\partial v^{y}_{\bold{k}}}{\partial k_x} - v^{x}_{\bold{k}} \frac{\partial v^{y}_{\bold{k}}}{\partial k_y} \right] \delta(E_{\bold{k}} - \mu )\nonumber\\
\sigma_{xx}^{T=0}(\mu) & =2e^2 \sum_{\pm} \int d \bold{k} \tau_{\bold{k}} (v^{x}_{\bold{k}})^2 \delta(E_{\bold{k}} - \mu ),\label{sigmaT0}
\end{align}

\begin{align}
\alpha_{xy}  & = \frac{1}{ e T }\int d\rho \frac{\partial f^{0}_{\rho}}{\partial \rho}(\rho-\mu)\sigma^{T = 0}_{xy}(\rho)\\
\alpha_{xx}  & = \frac{1}{e T}\int d\rho  \frac{\partial f^{0}_{\rho}}{\partial \rho}(\rho-\mu) \sigma^{T = 0}_{xx}(\rho) \\
\sigma_{xx} &= -\int d\rho   \frac{\partial f^{0}_{\rho}}{\partial \rho}\sigma^{T=0}_{xx}(\rho)\\
\sigma_{xy} &= -\int d\rho   \frac{\partial f^{0}_{\rho}}{\partial \rho}\sigma^{T=0}_{xy}(\rho).
\end{align}

In the limit where $\sigma^{T=0}_{ij}(\rho)$ varies slowly on the scale of $T$ such that $\partial_\rho \sigma^{T=0}_{ij}\gg T\partial_\rho^2 \sigma^{T=0}_{ij}$
the thermoelectric coefficients are given by the Mott relation
\begin{align}
\alpha_{ij} = - \frac{\pi^2}{3} \frac{T}{e} \frac{\partial \sigma^{T=0}_{ij}}{\partial \mu}.
\end{align}
In this low temperature limit, the expression for Nernst signal can be simplified as:
\begin{align}
\theta_{xy} = - \frac{\pi^2}{3} \frac{T}{e} \frac{\partial \Theta_{H}}{\partial \mu}\label{thetaxy}
\end{align}
where $\Theta_{H} = \frac{\sigma_{xy}^{T=0}}{\sigma^{T=0}_{xx}}$ is the Hall angle.
Since the Hall angle $\Theta_H$ changes on the scale of the Fermi energy, which is much larger than the temperature, the typical
contribution of a metal to the Nernst coefficient is small. This is referred to as the Sondheimer cancellation \cite{Ong}.

However, in the presence of a BDW order singular contributions to $\sigma_{ij}^{T=0}(\rho)$ can lead to sharp changes
in the Hall angle $\Theta_H$, that result in enhancement of the Nernst coefficient.
To compute the effects of the BDW with wave-vector $Q$ on $T=0$ conductivity (i.e. Eq.~\ref{sxx},~\ref{sxy}),
 it is convenient to change the integration variables from $k_{x}, k_{y}$ to $U = \frac{\epsilon_{\bold{k}} + \epsilon_{\bold{k}+\bold{Q}}}{2}$ and $V = \frac{\epsilon_{\bold{k}} - \epsilon_{\bold{k}+\bold{Q}}}{2}$.
The resulting conductivity is written as:
\begin{align}
\sigma^{T=0}_{xy}(\rho) & = \sum_{\pm} \int dU dV \frac{D(k_x,k_y)}{D(U,V)} \tau^2 v^{x} \left[ v^{y}  v^{xy} - v^{x} v^{yy} \right] \nonumber \\
& \times \delta(E - \rho )\label{eqsigma}\\
E(U,V) & = U + \alpha \sqrt{V^2 + W ^2} \qquad (\alpha = \pm 1)\label{sigmaxy}
\end{align}
where $v^{x}, v^{xy}$(detailed expression are given in the appendix) are band velocities that are related to derivatives of the energy $E$.

To understand the role of the BDW order parameter $W$, we focus on the limit of a small order parameter $W\ll |\mu|$.
 The contributions to the thermoelectric and conductivity response are dominated by ``hot spots" where the modification of Fermi surface by the BDW order is more dominant. 
In the small $W$ limit, deviations from the Sondheimer relations can be understood in terms of singular terms in the integrand of Eq.~\ref{sigmaT0}, which are written as:
\begin{align}
\frac{\partial E}{\partial V} & = \alpha \frac{V}{\sqrt{V^2+W^2}}\\
\frac{\partial^2 E}{\partial V^2} & = \alpha\frac{W^2}{(V^2 + W^2)^{\frac{3}{2}}}.
\end{align}
These terms develop singularity around the `hotspots' where $V\sim 0$ in the presence of a small BDW order parameter $W$.
The singularities at the hot-spot lead to
linear in $W$ contributions to Eq.~\ref{sigmaT0}. Since in this section we are interested in only linear $W$ contribution to Eq.~\ref{sigmaT0}, the other terms in Eq.~\ref{sigmaT0}  can be Taylor expanded at hot spots.

\begin{figure}
\includegraphics[scale=.3]{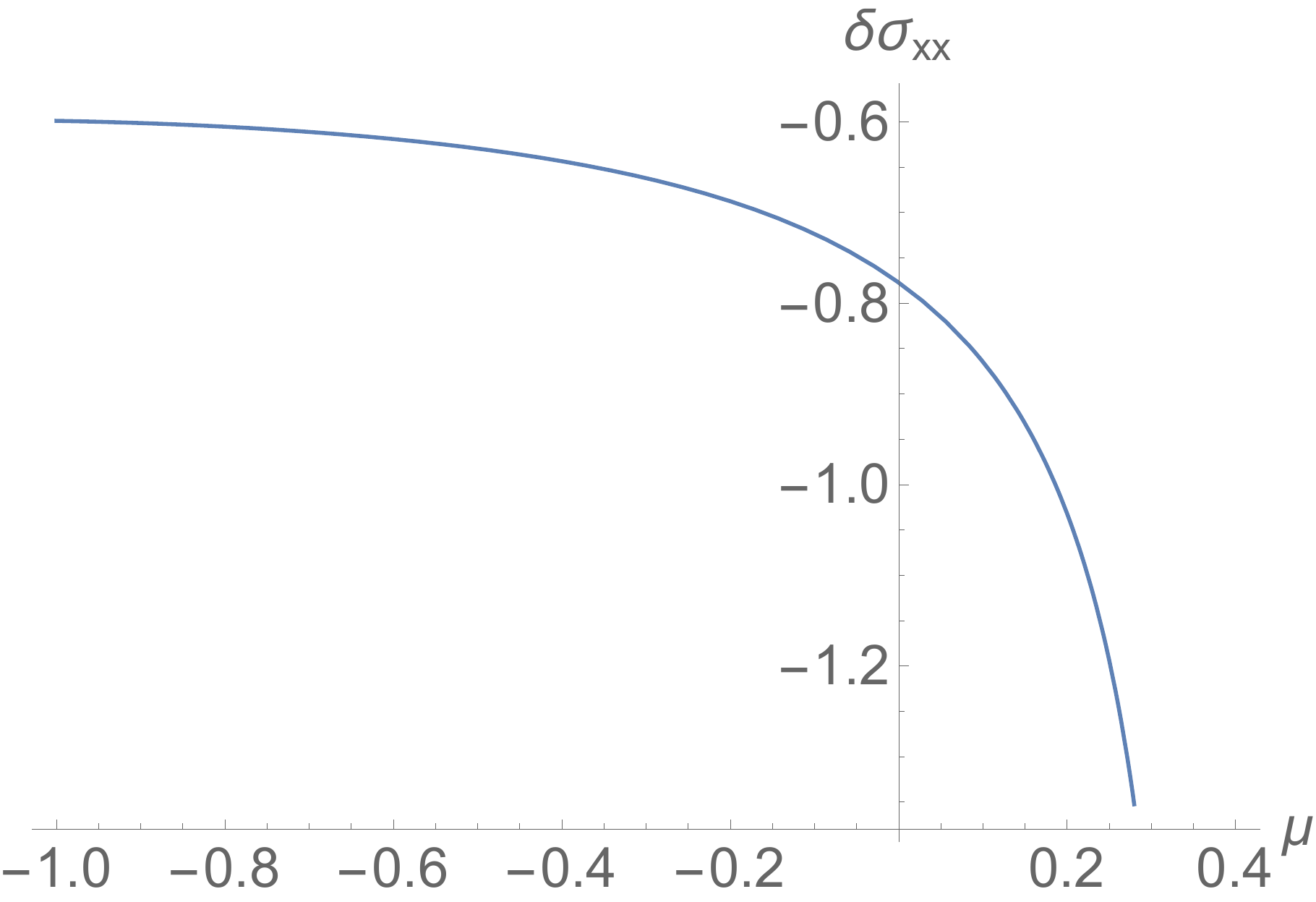}\\
\includegraphics[scale=.3]{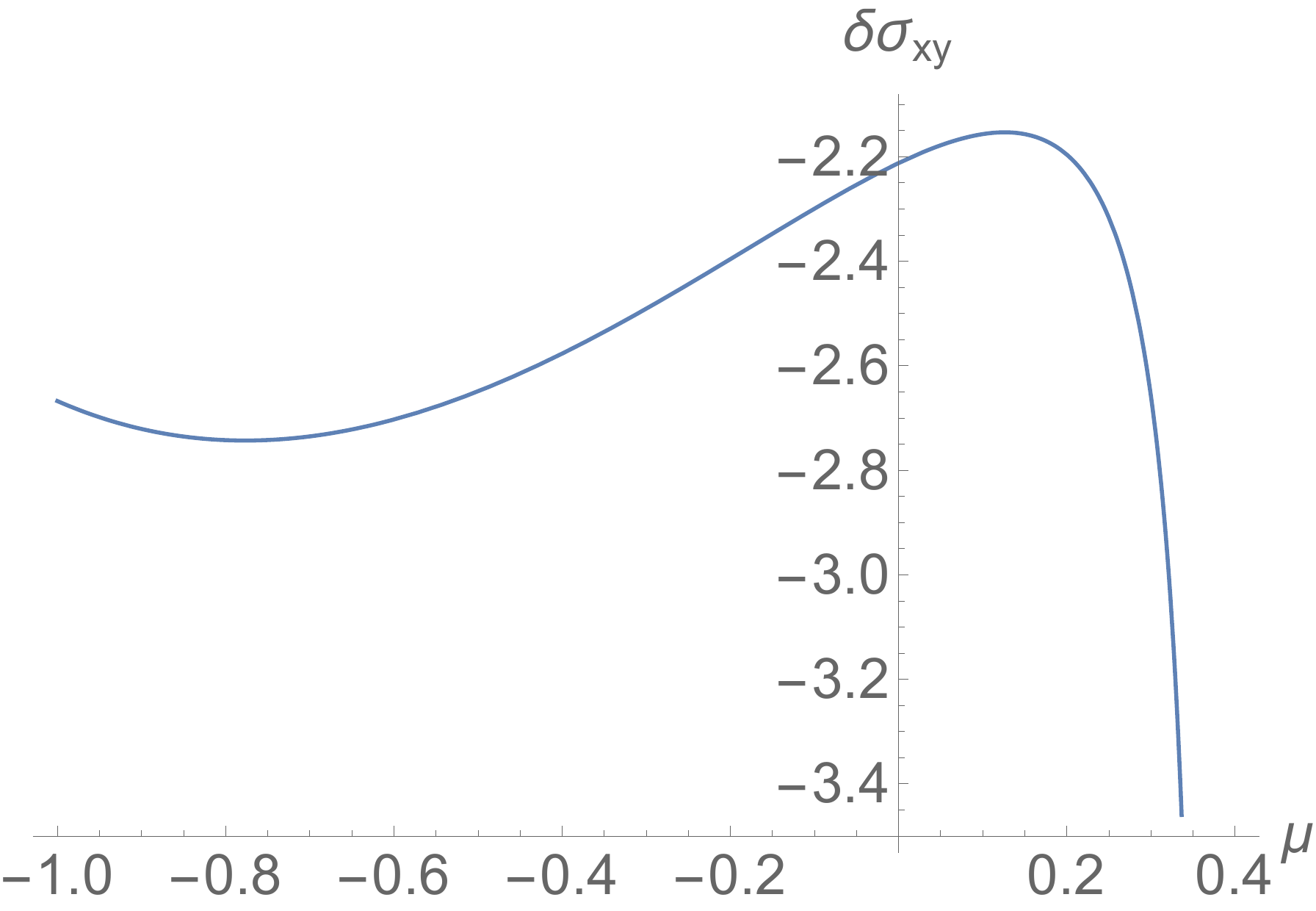}\\
\caption{The change of electrical conductivities as a function of chemical potential $\mu$. In this figure $-1 < \mu < -0.8 $ is the range of chemical potential corresponding to the underdoping.} \label{sigmafig}
\end{figure}

Defining the BDW induced correction to $\sigma^{T=0}$ as
\begin{align}
\delta \sigma_{ij}(\rho) = \sigma^{W,T=0}_{ij}(\rho) - \sigma^{W = 0,T=0}_{ij}(\rho)\label{sigmaij}
\end{align}
where $\sigma^{W,T=0}_{ij}(\rho)$ is the conductivity when there is BDW induced gap, while $\sigma^{W = 0,T=0}_{ij}(\rho)$ is the bare quantity.
The leading order for $\delta \sigma^{T=0}_{xy}(\rho)$ would be of $O(W)$, since results of $O(W^0)$ is subtracted by the unperturbed ones.
The non-singular contributions result in $O(W^2)$ contributions which we will ignore.
After simplifying Eq.~\ref{eqsigma} (details in the appendix), the leading order terms of $\delta \sigma_{xy}(\rho)$  is written as
\begin{align}
\delta \sigma_{xy}(\rho) = -2\pi(F_{1}+F_{2}+ G_{1}+G_{2})W \nonumber \\
+ \pi(F_{3} - F_{4}+G_{3} - G_{4}) W \label{deltasigmaxy}
\end{align}
where the expressions for $F$ and $G$ are involved and given in the appendix.
Similarly, we could obtain $\delta \sigma_{xx}(\rho)$:
\begin{align}
\delta \sigma_{xx}(\rho) = -2 \pi f_{1} W,
\end{align}
where as before the expression for $f_1$ is given in the appendix.
Expanding the conductivities $\sigma_{ij}$ for small $W$ according to Eq.~\ref{sigmaij} in Eq.~\ref{thetaxy}
we can write the linearized correction to the Nernst coefficient as 
\begin{align}
\delta \theta_{xy} & = - \frac{\pi^2}{3} \frac{T}{e} \frac{\partial }{\partial \mu} (\delta \Theta_{H})\Bigg |_{E_F} \\
& =  - \frac{\pi^2}{3} \frac{T}{e} \frac{\partial }{\partial \mu} \left( \frac{\delta \sigma_{xy} - \delta \sigma_{xx} \Theta^{0}_{H}}{\sigma^{0 2}_{xx}}\right)\Bigg |_{E_F}.\label{dthetaxy}
\end{align}
Using the parameters we set in the beginning of this paper, the numerical results for $\delta \sigma_{xx}$ and $\delta \sigma_{xy}$ in the range of chemical potential to our interest are shown in FIG. \ref{sigmafig}.
From the plots, we see the dominant contribution to the Nernst coefficient in the chemical potential range $-1<\mu<-0.7$ comes from   $\frac{\partial }{\partial \mu}(\delta \sigma_{xy})$ which leads to a positive Nernst signal. 
However, changes in parameters where the $\delta\sigma_{xy}$ curve shifts to lower chemical 
potentials will alter this sign to the experimentally consistent negative value. Such a shift can potentially be obtained by larger values of $W$ beyond linear response, which produces  a 
negative Nernst coefficient in a larger parameter range as we find in our numerical results described later. Another interesting feature of Fig.~\ref{sigmafig} is the divergence around $\mu\sim 0.3$. 
This leads to a dramatic enhancement of the Nernst coefficient, which will be the subject of the next section.

\section{Large Nernst signal for small $W$}
An interesting feature of the linear response results is a divergence in the conductance shifts $\delta\sigma$ in Fig.~\ref{sigmafig}. Since the Nernst 
 signal is proportional to the derivative of Hall angle over chemical potential, the divergence suggests a giant Nernst signal 
and a significant breakdown of the Sondheimer cancellation for specific structures of the fermi-surface.
In Fig.~\ref{sigmafig}, for a critical value of the chemical potential $\mu_c\sim 0.3$ both  $\delta \sigma_{xy}, \delta \sigma_{xx}$ appear to 
diverge. Examining Eq.~\ref{sigmaxy}, the  divergence can be viewed as a result of  the divergence in the Jacobian $\frac{D(k_x,k_y)}{D(U,V)}$
\begin{align}
\bold{J} = \frac{D(k_x,k_y)}{D(U,V)} = \frac{2}{|\bold{v}(\bold{k}+ \bold{Q}) \times \bold{v}(\bold{k}) |},\label{eq:J}
\end{align}
which appears in Eq.~\ref{sigmaT0}.

To derive the form of the divergence as a function of chemical potential $\mu$, we notice that for the $\bm{Q}=(Q,0)$ symmetry of the BDW,
\begin{align}
v_{x}(\bold{k}_0+ \bold{Q}) = -v_{x}(\bold{k}_0) \quad v_{y}(\bold{k}_0+ \bold{Q}) = v_{y}(\bold{k}_0)
\end{align}
where $\bold{k}_0$ is the position of hot spot.
Extracting the divergent terms in $\delta \sigma_{ij}$  we obtain:
\begin{align}
\delta \sigma_{xy} & = 4 \pi eB W_{0} \tau^2 \left( \frac{v_{x, \bold{k}_0+ \bold{Q}}}{v_{y,\bold{k}_0}} M^{-1}_{yy} \right)\\
\delta \sigma_{xx} & = -4 \pi W_{0} \tau \frac{v_{x, \bold{k}_0+ \bold{Q}}}{v_{y,\bold{k}_0}}
\end{align}
Substituting the above, we obtain:
\begin{align}
\delta \theta_{xy} = W_{0}\frac{4\pi^3 eB \tau^2}{3} \frac{T}{\sigma^{0}_{xx}} \frac{v_{x, \bold{k}_0+ \bold{Q}}}{v^{3}_{y,\bold{k}_0}}\\
\times \left[ 2 M^{-1}_{yy} + \Theta^{0}_{H}  \right] M^{-1}_{yy}
\end{align}
where $v_{y,\bold{k}_0} = \sqrt{2 M^{-1}_{yy} (\mu - \mu_c)}$, and $M_{yy}$ is the effective mass  at the hot spot. We see that there is a significant
 enhancement of Nernst signal around the critical chemical potential($\mu_c$). But the sign of the Nernst signal depends both on the bare Hall angle and 
the property of effective mass at hot spot and at critical chemical potential.

This mechanism produces a large Nernst signal by breaking down the Sondheimer cancellation when $\mu=\mu_c$ where Eq.~\ref{eq:J} diverges. This occurs when
the BDW wave-vector matches a nesting vector of the bare fermi surfaces. While the preceeding calculation is valid for small $W$ and explains the breakdown of Sondheimer cancellation at 
a chemical potential beyond the usually accepted range in the underdoped regime, it is conceivable that larger values of $W$ extend the range 
of breakdown of this cancellation. Besides, it has been proposed \cite{DChowdhury} that strong
correlation effects may lead to a low temperature electron pocket structure where the nesting would be similar to the required condition here i.e. $\mu\sim\mu_c$. 
Our result suggests that in addition
to providing a natural explanation for the BDW wave instability with the experimentally observed wave-vector direction, this correlated state \cite{DChowdhury}  would
 provide a mechanism for breakdown of Sondheimer cancellation even at small $W$.

\section{Numerical results for Nernst coefficient in BDW state}
\begin{figure}
\includegraphics[scale=.2]{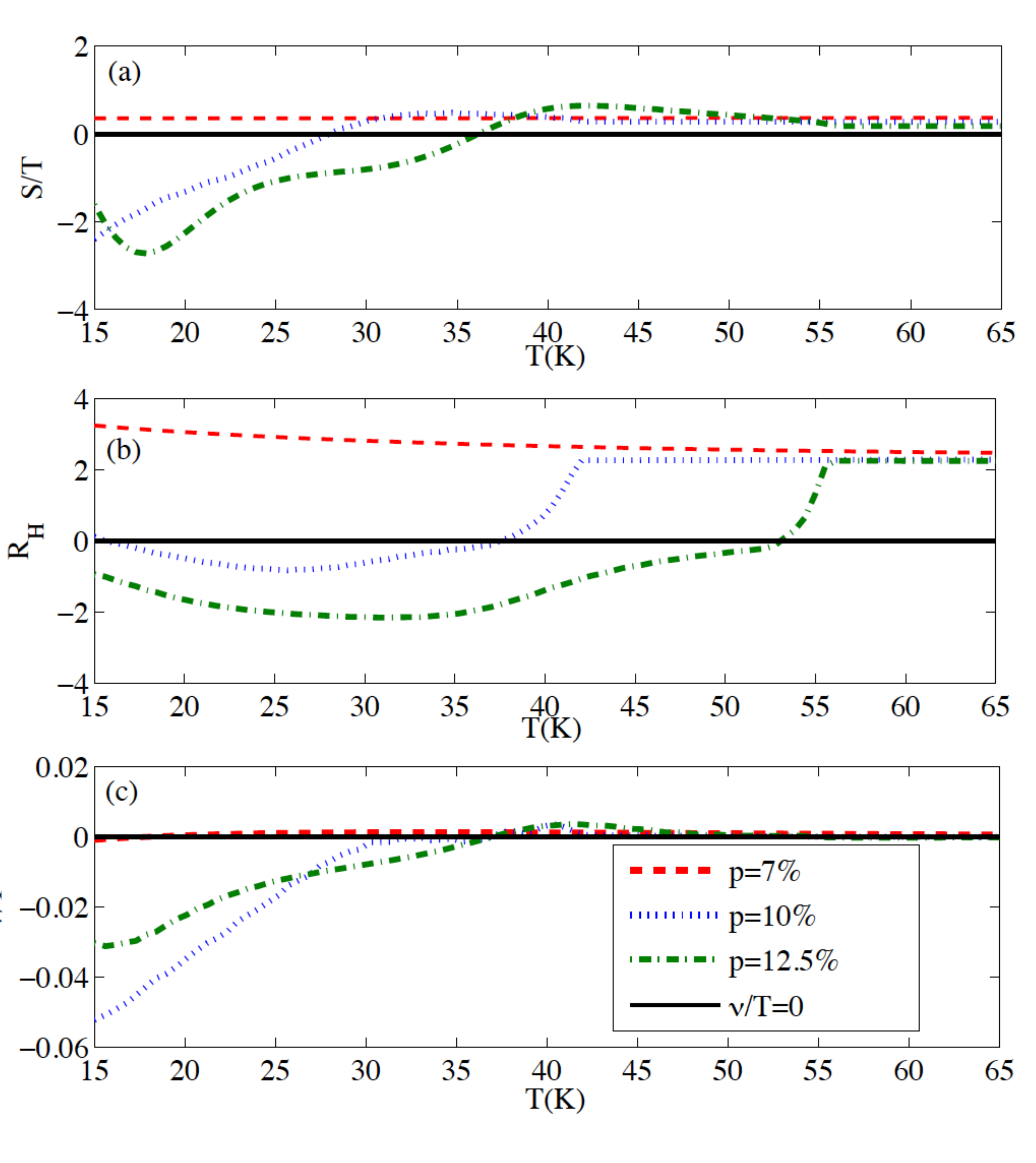}
\caption{(color online) Plot of transport coefficients vs. temperature in arbitrary units for three different doping values showing an enhanced negative signal in the underdoped regime (a) Seebeck coefficient $S/T$, (b) Hall coefficient $R_H$ and (c) Nernst coefficient $\nu/T$. The negative signals for doping values of $p=10\%$ and $p=12.5\%$ is ascribed to emergence of electron pockets due to Fermi surface reconstruction by BDW state. The Seebeck and Hall signal is positive for a doping of $p=7\%$ when $W=0$ while Nernst coefficient shows a small positive signal close to zero (when compared to the large negative signal for the other two doping values).}
\label{coeff_1}
\end{figure}
\begin{figure}
\includegraphics[scale=.2]{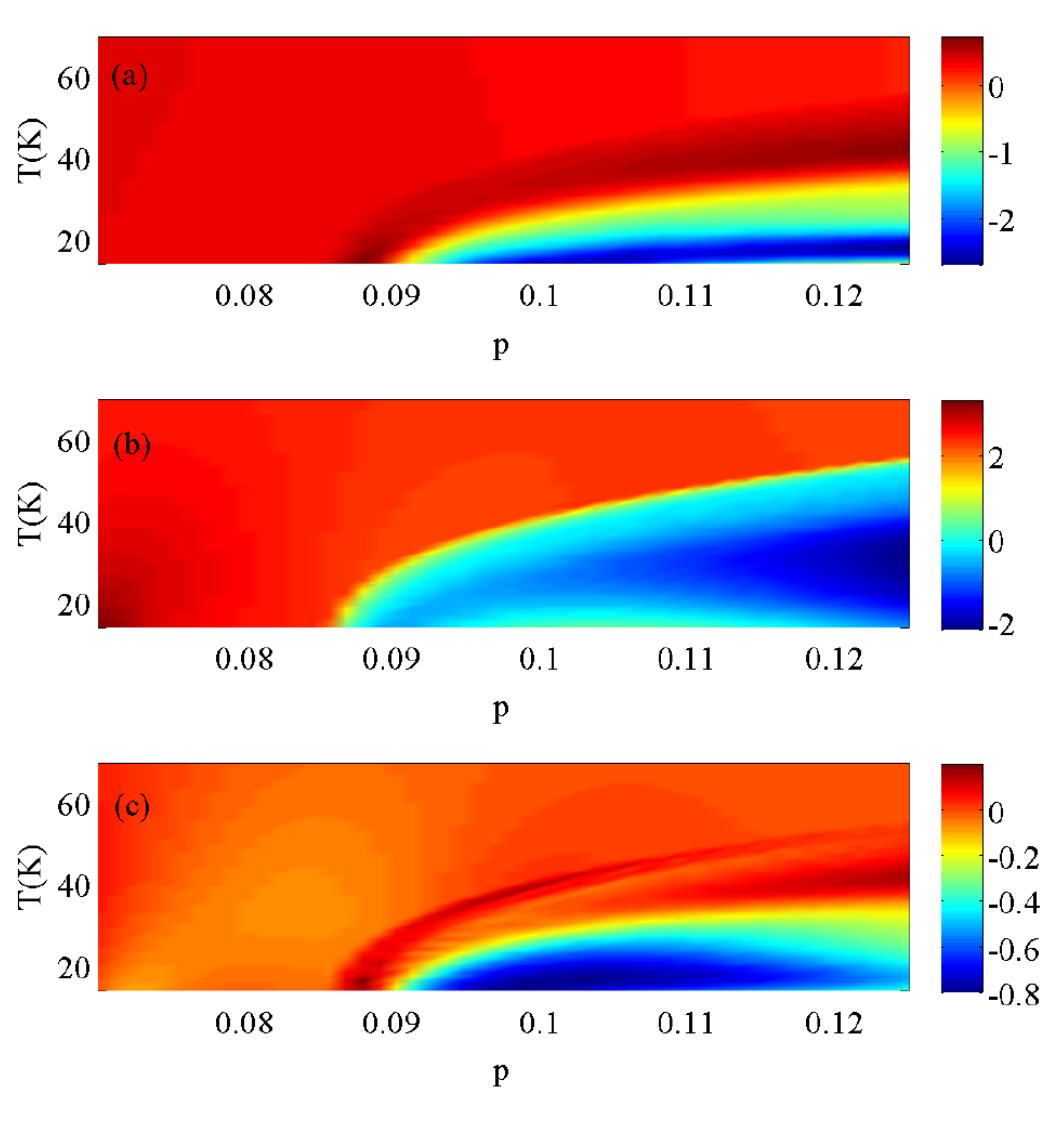}
\caption{(color online) Phase plot of transport coefficients in arbitrary units in the $p-T$ space showing an enhanced negative signal in the underdoped regime and low temperatures (a) Seebeck coefficient $S/T$, (b) Hall coefficient $R_H$ and (c) Nernst coefficient $\nu/T$. The dark blue area indicates the region of maximum negative response.}
\label{coeff_2}
\end{figure}
For numerical evaluation of the transport coefficients we choose values of the various parameters appropriate in the pseudogap phase. It is important to emphasize that our results (specifically the qualitative temperature and doping dependencies of the transport coefficients such as the Nernst coefficient) are completely robust against variations of the numerical values and functional forms of the various parameters. The functional dependence of the BDW order parameter on $T,p$ etc is chosen to qualitatively mimic the experimental trend, and is by no means meant to produce quantitatively accurate results for the Nernst and other transport coefficients in the pseudogap phase. We chose the bond density order parameter $W=0.22$ $eV$ at zero temperature and at doping $p=0.125$ (12.5\% or 1/8 hole doping) which is also set as the upper critical doping value $p_{up}$, for bond order to survive. For any finite value of doping below $p_{up}=0.125$, we assume a mean field doping dependence of $W(p)$ to be  $W|(p - p_{low})/(p_{up} - p_{low})|^{1/2}$, where we chose $p_{low}=0.085$ as the lower cutoff. For any other doping range $W(p)=0$. The bond order parameter is also assumed to scale mean field like with temperature below $T_{BDW}(p)$ as $W(p,T) = W(p)\sqrt{|1-T/T_{BDW}(p)|}$. From experimental fit, $T_{BDW}(p)$ is chosen to be $142$ $(p-p_l)^{0.3}$, which gives the critical temperature in Kelvins. For $T>T_{BDW}(p)$, $W(p,T)$ is again chosen to be zero.

Eq. \ref{axx} to \ref{sxy} were used to calculate normal state conductivities ($\alpha_{xx}$, $\alpha_{xy}$, $\sigma_{xx}$, $\sigma_{xy}$) at a give temperature $T$ and doping value $p$. Thus the $T$ dependence of the the Seebeck ($S=\alpha_{xx}/\sigma_{xx}$), Hall ($R_H=\sigma_{xy}/\sigma_{xx}\sigma_{yy}$) and Nernst coefficients ($\vartheta=(\alpha_{xy}\sigma_{xx} - \alpha_{xx}\sigma_{xy})/({\sigma_{xx}^2 + \sigma_{xy}^2})$) is calculated. Fig. \ref{coeff_1} shows normalized Seebeck and Nernst coefficients ($S/T$ and $\nu/T$) and Hall coefficient ($R_H$) as a function of temperature for three different doping values. For $p=10\%$ and $p=12.5\%$, we observe negative coefficients ascribed to electron like pockets due to BDW order but the signal remains positive for $p=7\%$ ($p<p_l$) doping when $W=0$. Fig. \ref{coeff_2} shows a phase space plot of Seebeck, Hall and Nernst coefficients in the $p-T$ phase space, where the region of enhanced negative response can be visualized to be in the pseudogap regime of low temperature and low doping.

\begin{comment}
\subsection{Uniaxial BDW}
It is worthwhile to examine if a unidirectional BDW modulation of $2\pi(0,1/3)$ or $2\pi(1/3,0)$ reproduces the negative response for the transport coefficients. We calculated Hall $R_H$ and Seebeck $S/T$ and Nernst $\nu/T$ coefficients obtained using Eq. \ref{axx} to \ref{sxy}. Fig. \ref{coeff_uni_1}  show the temperature and doping dependence for Hall and Seebeck coefficients. We find that this type of a unidirectional bond order does not produce a negative response signal. The Nernst coefficient also does not show an enhanced negative response when compared to bi-directional modulation.

\begin{figure}
\includegraphics[scale=.2]{coeff_uni_1_1.pdf}
\caption{Seebeck $(S/T)$ and Hall $(R_H)$ coefficients in arbitrary units obtained from bi-directional BDW order of type $\mathbf{Q}=(2\pi/3,0)$. Plots (a) and (b) show temperature dependence at three different doping values. Plots (c) and (d) show doping dependence for three different doping values. The signal remains positive throughout unlike a bi-directional BDW order.}
\label{coeff_uni_1}
\end{figure}
\end{comment}

\section{Conclusions}
In this work we studied the normal state of high $T_c$ cuprates i.e in the absence of superconductivity when a large magnetic field is applied. Starting with a mean field Hamiltonian for bi-directional BDW order with wave-vectors $\mathbf{Q}_1=(2\pi/3,0)$ and $\mathbf{Q}_2=(0,2\pi/3)$, we observed reconstruction of the Fermi surface from being large hole-like at higher doping (when there is no BDW order) to the appearance of small electron-like and hole-like pockets in the doping regime appropriate for the BDW state, which results from breaking of lattice translational symmetry. The normal state Nernst effect is important to understand the Fermi surface topology of cuprates in the underdoped regime. The enhancement and the negative sign of the low temperature Nernst signal experimentally observed in the pseudogap phase of cuprates is the main focus of this work. The Nernst response typically vanishes for conventional metals due to Sondheimer cancellation, but this cancellation breaks down in the presence of magnetic field at low temperatures due to the presence of a BDW order parameter. In addition to providing analytical understanding for the breakdown of the Sondheimer's cancellation in the presence of the BDW order parameter using the hot spot model, we numerically calculated all the three thermoelectric transport coefficients,  namely Hall, Seebeck and Nernst coefficients, in the semi-classical Boltzmann approximation. At the temperature scale $T<T_{BDW}$, we observed a negative Nernst coefficient in the underdoped regime. This low temperature negative Nernst response is reminiscent of a similar response of other two transport coefficients, namely the Hall and Seebeck coefficients. Though the negative sign of the Hall and Seebeck coefficients can be ascribed to the appearance of electron-like pockets which appear on the Fermi surface, the sign of the Nernst coefficient is not directly determined by the sign of the dominant charge carriers and depends on the detailed Fermi surface topology.  

Acknowledgment: G.S and S.T are supported by AFOSR
(FA9550-13-1-0045). C.L and J.D.S would like to acknowledge the
University of Maryland, Condensed Matter theory center, and
the Joint Quantum institute for startup support.
\begin{comment}
We also studied the Fermi surface topology and the transport coefficients for a uni-axial BDW of type $\mathbf{Q}_1$ or $\mathbf{Q}_2$. This type of a bond order modulation does not reproduce the results of the bi-directional modulation and is thus inconsistent with experiments. The sign of the coefficients remained positive which we ascribe to the hole-like pocket in the Fermi surface.
\end{comment}

\appendix

\section{Expansion of Nernst coefficient}

The explicit expression for band velocity $v_{\alpha}$ is
\begin{align}
v^x & = \frac{\partial E}{\partial k_x} =  \frac{\partial U}{\partial k_x}\frac{\partial E}{\partial U} + \frac{\partial V}{\partial k_x}\frac{\partial E}{\partial V} \nonumber \\
& = U_{x} + V_{x}\left( \frac{\partial E}{\partial V}\right) \label{vx}\\
v^y & = U_{y} + V_{y}\left( \frac{\partial E}{\partial V}\right)
\end{align}

The inverse of effective mass terms are:
\begin{align}
v^{xx} & =  \frac{\partial v^{x}}{\partial k_x} = U_{xx} + V_{xx}\left( \frac{\partial E}{\partial V}\right) + (V_{x})^2 \frac{\partial ^2 E}{\partial V^2}\\
v^{yy} & =  \frac{\partial v^{y}}{\partial k_y} = U_{yy} + V_{yy}\left( \frac{\partial E}{\partial V}\right) + (V_{y})^2 \frac{\partial ^2 E}{\partial V^2}\\
v^{xy} & =  \frac{\partial v^{x}}{\partial k_y} = U_{xy} + V_{xy}\left( \frac{\partial E}{\partial V}\right) + V_{x}V_{y} \frac{\partial ^2 E}{\partial V^2},
\end{align}

where

\begin{align}
E & = U \pm \sqrt{V^2+ W ^2}\\
U_{x} & = \frac{\partial U}{\partial k_{x}} \quad U_{y}  = \frac{\partial U}{\partial k_{y}}\\
V_{x} & = \frac{\partial V}{\partial k_{x}} \quad V_{y}  = \frac{\partial V}{\partial k_{y}}\\
U_{xx} & = \frac{\partial U}{\partial k_x} \left( \frac{\partial U_{x}}{\partial U}\right) + \frac{\partial V}{\partial k_x} \left( \frac{\partial U_{x}}{\partial V}\right) \\
U_{xy} & = \frac{\partial U}{\partial k_y} \left( \frac{\partial U_{x}}{\partial U}\right) + \frac{\partial V}{\partial k_y} \left( \frac{\partial U_{x}}{\partial V}\right) \\
U_{yy} & = \frac{\partial U}{\partial k_y} \left( \frac{\partial U_{y}}{\partial U}\right) + \frac{\partial V}{\partial k_y} \left( \frac{\partial U_{y}}{\partial V}\right) \label{uyy}
\end{align}

While calculating the leading order of $ \delta \sigma_{xy}^{T=0}(\rho)$, we substitute in the expressions between eqs.~\ref{vx} and \ref{uyy} into Eq.~\ref{sigmaT0}, and perform the  integration, and keep only the linear order terms in $W$.  The result is Eq.~\ref{deltasigmaxy}, where, 
\begin{align}
F_{1}(\rho,0) & = \tau^2 J(U,V) \left( U_{x}V_{y}V_{xy} +  V_{x}U_{y}V_{xy} + V_{x}V_{y}U_{xy} \right) |_{(\rho,0)}\\
F_{2}(\rho,0) &= \frac{\partial}{\partial U}\left( \tau^2 J U_{x}U_{y}V_{x}V_{y}\right) |_{(\rho,0)}\\
F_{3}(\rho, 0) &= \frac{\partial}{\partial V} \left[ \tau^2 J \left( U_{x}V_{y}V_{x}V_{y} + V_{x}U_{y}V_{x}V_{y} \right) \right] |_{(\rho,0)}\\
F_{4}(\rho,0) & = \frac{\partial}{\partial U}\left[ \tau^2 J\left( V_{x}V_{y}V_{x}V_{y} \right)\right]|_{(\rho,0)}
\end{align}

Here, we have assumed  the integral limits of $V$ to be $V/ W \rightarrow \pm \infty$. The above expression only represent the first part in the square bracket of the first equation in Eq.~\ref{sigmaT0}. For the second part, we just change the derivative variable from $xyxy$ to $xxyy$, a minus in the front and thus obtain $G$'s. When we consider $\bold{Q}_y$, because of the $C_4$ rotational symmetry, we simply interchange $x$ and $y $.

Similarly, we get the expression for $\delta \sigma_{xx}$ in Eq.~\ref{sigmaT0}, where
\begin{align}
f_{1}(\rho,0) = \tau J (V_{x})^2 |_{(\rho,0)}
\end{align}

\section{Hamiltonian matrix elements}
The non zero elements of the 9 component Hamiltonian are specifically given by:
\begin{widetext}
\begin{eqnarray*}
w_{12}&=W_0\left(\cos\left(k_x+{\pi}/{3}\right)-\cos k_y\right)\\
w_{13}&=W_0\left(\cos\left(k_x-{\pi}/{3}\right)-\cos k_y\right)\\
w_{14}&=W_0\left(\cos k_x-\cos\left(k_y+{\pi}/{3}\right)\right)\\
w_{17}&=W_0\left(\cos k_x-\cos\left(k_y-{\pi}/{3}\right)\right)\\
w_{23}&=W_0\left(\cos\left(k_x+\pi\right)-\cos k_y\right)\\
w_{25}&=W_0\left(\cos\left(k_x+{2\pi}/{3}\right)-\cos \left(k_y+{\pi}/{3}\right)\right)\\
w_{28}&=W_0\left(\cos\left(k_x+{2\pi}/{3}\right)-\cos \left(k_y-{\pi}/{3}\right)\right)\\
w_{36}&=W_0\left(\cos\left(k_x-{2\pi}/{3}\right)-\cos \left(k_y+{\pi}/{3}\right)\right)\\
w_{39}&=W_0\left(\cos\left(k_x-{2\pi}/{3}\right)-\cos \left(k_y-{\pi}/{3}\right)\right)\\
w_{45}&=W_0\left(\cos\left(k_x+{\pi}/{3}\right)-\cos \left(k_y+{2\pi}/{3}\right)\right)\\
w_{46}&=W_0\left(\cos\left(k_x-{\pi}/{3}\right)-\cos \left(k_y+{2\pi}/{3}\right)\right)\\
w_{47}&=W_0\left(\cos k_x-\cos \left(k_y+\pi\right)\right)\\
w_{56}&=W_0\left(\cos\left(k_x+\pi\right)-\cos \left(k_y+{2\pi}/{3}\right)\right)\\
w_{58}&=W_0\left(\cos\left(k_x+{2\pi}/{3}\right)-\cos (k_y+\pi)\right)\\
w_{69}&=W_0\left(\cos\left(k_x-{2\pi}/{3}\right)-\cos (k_y+\pi)\right)\\
w_{78}&=W_0\left(\cos\left(k_x+{\pi}/{3}\right)-\cos \left(k_y-{2\pi}/{3}\right)\right)\\
w_{79}&=W_0\left(\cos\left(k_x-{\pi}/{3}\right)-\cos \left(k_y-{2\pi}/{3}\right)\right)\\
w_{89}&=W_0\left(\cos\left(k_x+\pi\right)-\cos \left(k_y-{2\pi}/{3}\right)\right)\\
\end{eqnarray*}\end{widetext}

\end{document}